%% file: main.tex
\documentclass[runningheads]{llncs}

\input{packages.tex}

\title{Probabilistic Quality of Service aware Service Selection
	\thanks{Research partly supported by the European Unions Horizon 2020 research and innovation programme under the Marie Skodowska-Curie grant agreement No 778233. Carlos G.\ Lopez Pombo's research is supported by Universidad de Buenos Aires through grant UBACyT 20020170100544BA, and Consejo Nacional de Investigaciones Científicas y Técnicas through grant PIP 11220130100148CO.}
}

\author{Agustín E. Martinez Suñé\inst{1,2} \and Carlos G. Lopez Pombo\inst{1,2}\thanks{Now on leave from the Universidad de Buenos Aires to the Escuela de Producción, Tecnología y Medio Ambiente, Universidad Nacional de Río Negro.}}

\authorrunning{A.E. Martinez Suñé et. al.}
\institute{Universidad de Buenos Aires. Departamento de Computación. \and CONICET--UBA. Instituto de Investigación en Ciencias de la Computación (ICC).
} 
\index{Lopez Pombo, Carlos G.}
\index{Martinez Suñé, Agustín E.}

\input{macros}

\begin{document}

\maketitle

\begin{abstract}
In software-as-a-service paradigms software systems are no longer monolithic pieces of code executing within the boundaries of an organisation, on the contrary, they are conceived as a dynamically changing collection of services, collectively executing, in pursuit of a common business goal.
An essential aspect of service selection is determining whether the \emph{Quality of Service} (QoS) profile of a service satisfies the QoS requirements of a client.

In realistic execution environments, such QoS values might be influenced by external, non-controllable events, making it impossible for the service provider to guarantee that the values characterised by a QoS profile will be met, naturally leading to the need of a probabilistic interpretation of QoS profile. 

In this work we propose:
\begin{inparaenum}[1)]
\item a model for describing probabilistic QoS profiles based on multivariate continuous probability distributions,
\item a language for describing probabilistic QoS requirements, and
\item an automatic procedure for assessing whether a probabilistic QoS profile satisfies a probabilistic QoS requirement.
\end{inparaenum}
\end{abstract}

\input{introduction}

\input{probabilisticqos}
\input{learning}
\input{conclusions}

\bibliography{zoterolib,morebib}
\bibliographystyle{splncs}

\end{document}

%% file: packages.tex
\usepackage[table,x11names]{xcolor}
\usepackage[english]{babel}
\usepackage[utf8]{inputenc}
\usepackage{todonotes}
\usepackage{paralist}
\usepackage{amssymb}
\usepackage{amsfonts}
\usepackage{amsmath}
\usepackage{wrapfig}
\usepackage{multirow}

\usepackage[linesnumbered]{algorithm2e}
\let\oldnl\nl
\newcommand\nonl{%
\renewcommand{\nl}{\let\nl\oldnl}}

\usepackage{cleveref}

\usepackage[shortlabels]{enumitem}
\newlist{todolist}{itemize}{2}
\setlist[todolist]{label=\huge$\square$}
\usepackage{pifont}
\usepackage{ulem}

\usepackage{float}

%% file: macros.tex
\newcommand{\randvec}{\mathbf{X}}
\newcommand{\randvar}{X}
\newcommand{\randobs}{x}
\newcommand{\vect}[1][]{
    \ifthenelse{\equal{#1}{}}{\mathbf{x}}{\mathbf{x_#1}}
}

\newcommand{\qosprf}{\randvec}

\newcommand{\qoscstr}{\varphi}
\newcommand{\qosreq}{\Phi}
\newcommand{\qosattr}{a}
\newcommand{\qosattrset}{A}

\newcommand{\qoscstrset}{\mathit{QoSCstr}}
\newcommand{\qosreqset}{\mathit{QoSReq}}

\newcommand{\region}{R}
\newcommand{\pmin}{{p_\mathit{min}}}
\newcommand{\pmax}{{p_\mathit{max}}}

\newcommand{\Prob}{\mathbb{P}}

\newcommand{\propvar}{p}
\newcommand{\propvarb}{c}
\newcommand{\propval}{v}
\newcommand{\propset}{\mathcal{P}}

\newcommand\Set[2]{\{\,#1\mid#2\,\}}

%% file: introduction.tex

\section{Introduction}
\label{sec:intro}

In software-as-a-service paradigms such as \emph{Service-oriented Computing} -- SOC and Cloud/Fog/Edge computing, software systems are no longer monolithic pieces of code executing within the boundaries of an organisation. On the contrary, this new generation of applications run over globally available computational resources and communication infrastructure. They rely on the intervention of a dedicated middleware responsible for the discovery of services that are bound at runtime, subject to the negotiation of a \emph{Service Level Agreement} -- SLA, so they can collectively fulfill a given business goal \cite{fiadeiro:fac-23_4}.
From this viewpoint, complex applications are conceived as the composition of publicly available software services that act as fundamental building blocks \cite{bouguettaya14}.

The emergence of this paradigm has been accompanied by a deep transformation of the business models associated with the construction of software systems. Providers of cloud computing platforms rely on these notions and offer a high degree of customisation for their services with which companies can configure the resources to better suit their business needs. Well-known examples of such a customisation are pricing schemes that depend on the amount of time a computational resource is used. Emerging technologies such as \emph{serverless computing} or \emph{Function as a Service} -- FaaS increase the possibilities for service-based paradigms, as multiple providers offering services might compete to be selected by software systems that take the role of service consumers. An essential aspect of such a service selection (known as the Service Selection Problem \cite[pt. II]{bouguettaya14}) is determining whether the \emph{Quality of Service} (QoS) profile of a service (i.e., a description of the potential values that its quantitative attributes might adopt) satisfies the QoS requirements of a client. The interested reader is pointed to \cite{moghaddam14} for a comparative review of existing approaches and to \cite{hayyolalam:jnca-110} for a systematic literature review of the problem. 

The QoS profile of a service can be thought of as a specifications describing the values adopted by its quantitative attributes (i.e., attributes that admit some type of measurement, for example, \emph{response time}, \emph{memory usage}, \emph{price}, \emph{reputation}, etc.) along its possible executions.  
In realistic execution scenarios such values are usually subject to external, non-controllable events like hardware failures, communication errors, congestion in the communication network, etc. Therefore, it is utterly difficult, if not impossible, to guarantee that the values characterised by a QoS profile will always be met with absolute certainty, naturally leading to a probabilistic interpretation of the QoS profile, where an execution yields a valuation of the quantitative attributes with certain probability. Similarly, we need a probabilistic interpretation of the QoS requirements; in it, a client will bind to a specific service provider only if the probability of its quantitative attributes adopting certain values is within prespecified bounds. 

\subsection{Our approach}
\begin{figure}[H]
    \centering
    \includegraphics[width=.75\textwidth]{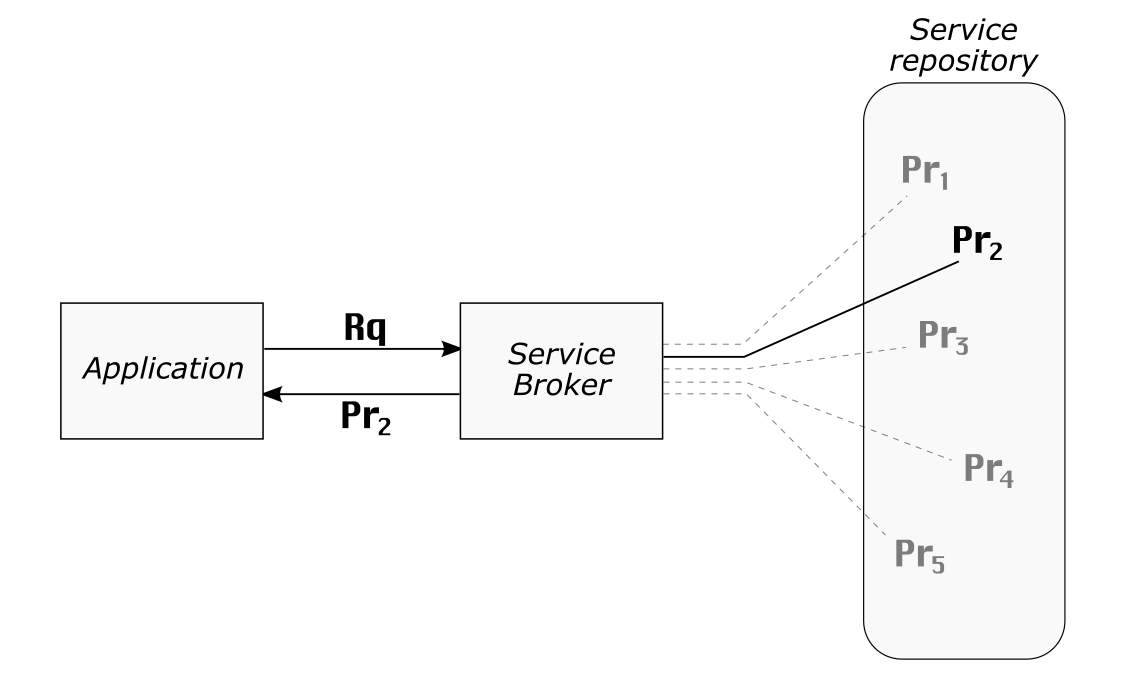}
    \caption{A schematic view of the QoS-aware service selection problem.}
    \label{fig:service-selecion-problem}
\end{figure}

Any approach on this problem is composed of a specific view in each of the following four dimensions: 
\begin{inparaenum}[1)]
    \item the precise definition of the \emph{problem}, 
    \item the \emph{QoS profile model} for describing each service, 
    \item the \emph{QoS requirement model} for describing the client's needs, and 
    \item the \emph{decision procedure} for automatically selecting the appropriate service.    
\end{inparaenum}
In the following, we outline this dimensions for our approach and we compare it with existing approaches in the literature.
\paragraph{Problem} Automatic probabilistic QoS-aware service selection: given a QoS requirement of a client running application and a set of QoS profiles for functionally equivalent services, choose a service whose QoS profile satisfies the QoS requirement. In \cref{fig:service-selecion-problem} we show a schematic view of this problem where a \emph{Service Broker} plays the role of on-demand choosing a \emph{Service Provider} from a \emph{Service repository} in order to satisfy the needs of an application, playing the role of \emph{Service Client}.
\paragraph{QoS profile model} The QoS profile of each service is modeled as a multivariate continuous probability distribution where each random variable denotes a specific QoS attribute. We propose a methodology for learning the QoS profile of a service from data sampled from the real-world behaviour of such service.
\paragraph{QoS requirement} QoS requirements are given by the client application to drive the service selection. They are boolean combinations of QoS constraints which, in turn, specify probability bounds for the QoS attributes taking values within a polyhedral region.
\paragraph{Decision procedure} Given a QoS profile model and a QoS requirement formula our algorithm determines if the QoS profile satisfies the requirement. It leverages on SAT-solving to handle the boolean structure of the requirement formula and on Monte Carlo integration to determine if the probability bounds prescribed by the QoS constraints hold.

All this done in a general setting where there is no \emph{a priori} assumption about the specific nature of the quantitative attributes (i.e. any quantifiable characteristic of software service can be considered a quantitative attribute).


\subsection{Related work}
\paragraph*{Problem} There is an extensive body of work focusing on the Quality of Service aspect of service selection and service composition in a cloud environment \cite{hayyolalam:jnca-110}. The precise definition of the problem setting varies from work to work. While some works consider the selection of serveral services to compose according to a given workflow \cite{zheng:atw-10_2}, others conceive the problem as the selection of one service among a given set of services \cite{yau:icsc11}. In this work we adopt the latter notion, we understand it as a stepping stone for putting forward our probabilistic approach, which is a necessary building block for tackling the service composition problem in a probabilistic setting.
\paragraph*{QoS profile model}
The models proposed for characterizing QoS attributes can be grouped into three broad categories: single-value attributes, range-valued attributes, and probabilistic attributes. In the single-value setting the QoS provided for a particular attribute is modeled as a constant value \cite{liangzhaozeng:itse04,yau:icsc11} while in the range-valued setting it is usually modeled as an interval or a set of values \cite[Section~2]{martin-diaz:03}. The latter perspective emerges by acknowledging the limitations of the former: there are scenarios in which the single value model becomes insufficient. 
Instead, we position our work as part of an increasing number of research interested in capturing the probabilistic nature of the QoS attributes \cite{klein:icsoc09,rosario:icws09,hwang:is07,hwang:itsc-8_3}. In \cite{klein:icsoc09,rosario:icws09} the authors propose to solve the problem of probabilistic QoS based service selection, but they do it in a setting where QoS attributes are modeled as separate probability distributions that are independent. In \cite{hwang:is07} the authors propose a probabilistic model for describing QoS but only considering discrete probability distributions, also restricting attributes to five specific metrics. In \cite{hwang:itsc-8_3} the authors address the service selection problem with probabilistic models, but each QoS attribute is modeled as a discrete random variable with a probability mass function, leaving out continuous probability models.

\paragraph*{QoS requirement}
Most of the contributions to QoS modeling for the service selection problem focus on the description of the QoS profile, as stated in the last paragraph. In such contributions, the notion of QoS requirement is relagated as an implicit optimization criteria \cite[Section~5.1]{moghaddam14}, where the aim is to minimize or maximize the value of each QoS attribute according to the nature of the attribute. Instead, we consider QoS requirements to be a key elements since it allows the service selection to be specific to the clients needs. Under this view, the proposal of a language for describing QoS requirements becomes relevant.
\paragraph*{}
Other distinctive contributions can be found in the field of probabilistic quantitative verification of service-based systems \cite{calinescu:ieeetse-37_3}. There, a more classical verification-like framework is proposed where Markov models are used to determine the validity of probabilistic temporal logic formulae representing quality of service properties. In \cite{calinescu:ieeetr-65_1} the general approach is enriched with a relativised notion of validity within a confidence interval. The main difference between these approaches and ours relies on the model of the system being analyzed. While probabilistic model checking approaches need a state-based model of the behaviour of the system, our approach only needs a model of the values taken by the QoS attributes of such system. This is a consequence of approaching the problem from a service oriented computing perspective, where services are considered to be opaque entities from the perspective of the consumer (i.e., the consumer has no knowledge of their internal behaviour).

We also acknowledge the existence of an important body of knowledge that focuses on the problem of choosing an optimum candidate according to prefixed preferences based on soft constraint solving \cite{zemni:icsoc10,rosenberg:edoc09}, 
and also using fuzzy set theory to express preferences \cite{chouiref:asc-41}, but we will concentrate on a more classical, thus strict, notion of satisfaction of QoS requirements, which state probabilistic requirements imposed to service providers.

There are three key ingredients in our approach and, to the best of our knowledge, ours is the first proposal to tackle all of these aspects simultaneously:
\begin{inparaenum}[1)]
\item we propose a modeling language capable of dealing with an unbounded number QoS attributes of which there is no implicit interpretation (i.e., we only require that they can be measured in each concrete execution of the service),
\item we focus on continuous probability distribution, capable of capturing a finer grain characterisation of the services behaviour, and 
\item (probably the most novel ingredient) we propose the use of multivariate probability distribution, enabling the characterisation of dependencies between the probabilistic behaviour of the QoS attribute.
\end{inparaenum}

The paper is organised as follows.
\begin{inparaenum}[]
\item In \cref{sec:probqossla} we present the definitions of QoS profile and QoS requirements, together with a satisfaction relation; we also frame the service selection problem in terms of these three elements.
\item In \cref{sec:implementation} we propose an algorithm implementing the satisfaction relation proposed in \cref{sec:probqossla}.
\item In \cref{sec:learning} we propose a methodology for learning probabilistic QoS profiles from data sampled from the real-world behaviour of the services, taking out the responsibility of producing it from the service provider. 
\item Finally, in \cref{sec:conclusions} we draw some concluding remarks and propose some further research directions. 
\end{inparaenum}

%% file: probabilisticqos.tex

\section{Probabilistic QoS-aware SLA}
\label{sec:probqossla}
In this section we present the main contributions of this work. We start by providing a probabilistic definition of QoS profiles for service providers and QoS requirements for service clients, together with a definition of the satisfaction relation between QoS profiles and QoS requirements.


\subsection{Probabilistic QoS profile of a service}\label{sec:qosprofile}
Let us start by discussing how the probabilistic QoS behaviour of a software service is to be modeled. We focus on \emph{quantitative attributes} of software artifacts (i.e., those admitting some type of measurement), such as \emph{response time}, \emph{throughput}, \emph{price}, \emph{reputation}, among others; and, in our case, not assuming any additional property, interpretation or behaviour. The sole assumption is that the attributes can be actively monitored, at each execution of the service, by an element of the execution infrastructure, for example the \emph{Service Broker}.

\begin{figure}[h]
    \centering
    \includegraphics[width=.45\textwidth]{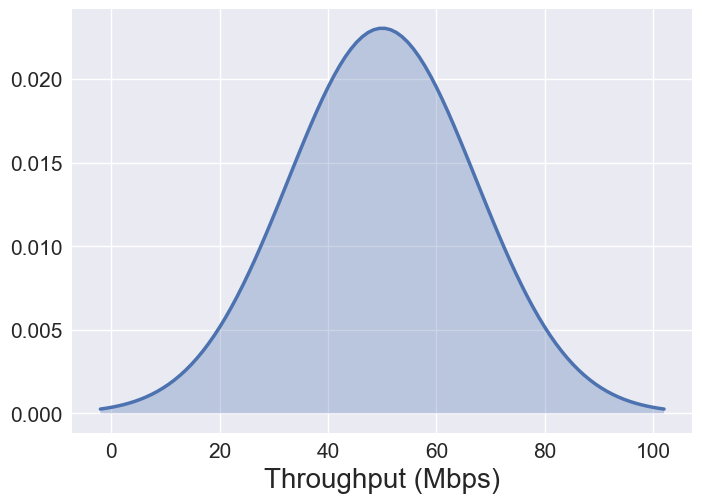}
    \includegraphics[width=.46\textwidth]{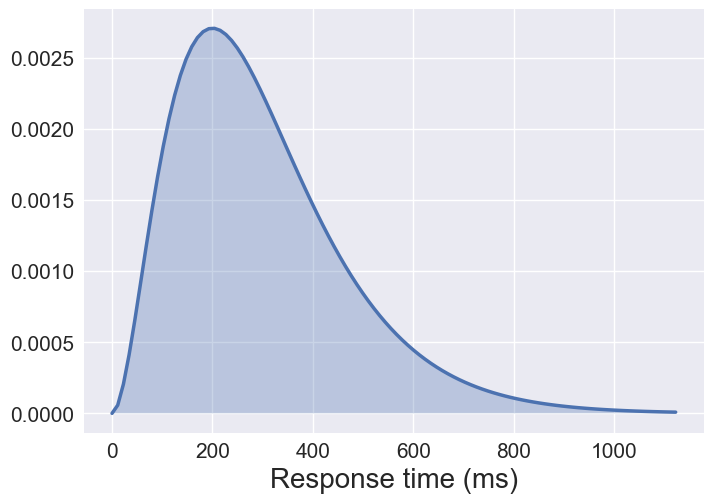}
    \caption{Probabilistic QoS profile where TP follows a Gaussian distribution with $\mu = 50$ and $\sigma^2 = 300$ and RT follows a Gamma distribution with $\alpha = 3$ and $\beta = \frac{1}{100}$.}
    \label{fig:independent-example}
\end{figure}

As a running example, consider a set of (functionally equivalent) services for validating an email address; an example taken from \cite{al-masri:icccn07}. Upon calling one of these services, the client provides an email address and the service indicates whether it is a valid and existing email. Different services have different values of throughput (TP) and response time (RT). As we are interested in the probabilistic nature of quantitative attributes, given a candidate service, we would like to be able to capture a situation like the one shown in \cref{fig:independent-example}, where the TP value offered by a given service is known to follow a \emph{Gaussian} distribution with \emph{mean} in 50 Mbps, and the RT value is known to follow a \emph{Gamma} distribution with mean in 300 ms.

From now on, definitions and results will be presented assuming a fixed, but arbitrary, set of quantitative attributes $\qosattrset = \{\qosattr_1, \ldots, \qosattr_n\}$.

\begin{definition}[QoS profile]
A \emph{QoS profile} is a random vector $\randvec = (\randvar_1, \ldots, \randvar_n)$, where each random variable $\randvar_i$ is interpreted as the value of the quantitative attribute $\qosattr_i$, and is characterised by a joint cumulative distribution function (CDF): $F_\randvec(\randobs_1, \ldots, \randobs_n) = \Prob(\randvar_1 \leq \randobs_1, \ldots, \randvar_n \leq \randobs_n)$

Moreover, if the joint CDF can be differentiated, then $\randvec$ is characterised by a joint Probability Density Function (PDF): $f_\randvec(\randobs_1, \ldots, \randobs_n) = \frac{\partial^n F(\randobs_1, \ldots, \randobs_n)}{\partial_{\randobs_1} \ldots \partial_{\randobs_n}}$
\end{definition}

Considering the attributes TP and RT, mentioned in \cref{fig:independent-example}, the QoS profile is the random vector $\randvec = (\randvar_1, \randvar_2)$, where $\randvar_1$ and $\randvar_2$ represent the value of TP and  RT, respectively. Then, the PDF is defined as $f_{\randvec}(\randobs_1, \randobs_2) = f_{\randvar_1}(\randobs_1) \cdot f_{\randvar_2}(\randobs_2)$, where $f_{\randvar_1}(\randobs_1) = \frac{1}{\sigma\sqrt{2\pi}} \exp[-\frac{1}{2}\left(\frac{\randobs_1 - \mu}{\sigma}\right)^2]$ with $\mu = 50$, $\sigma^2 = 300$, and  $f_{\randvar_2}(\randobs_2) = \frac{\beta^\alpha}{\Gamma(\alpha)} {\randobs_2}^{\alpha - 1} \exp(-\beta \randobs_2)$ with $\alpha = 3 \text{ and } \beta = \frac{1}{100}$.

\begin{figure}[h]
    \centering
    \includegraphics[width=.45\textwidth]{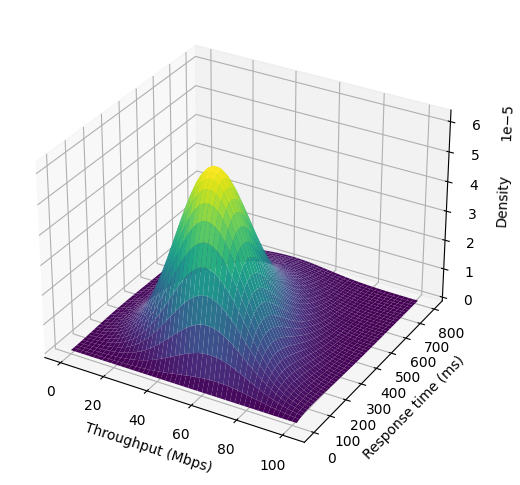}
    \includegraphics[width=.45\textwidth]{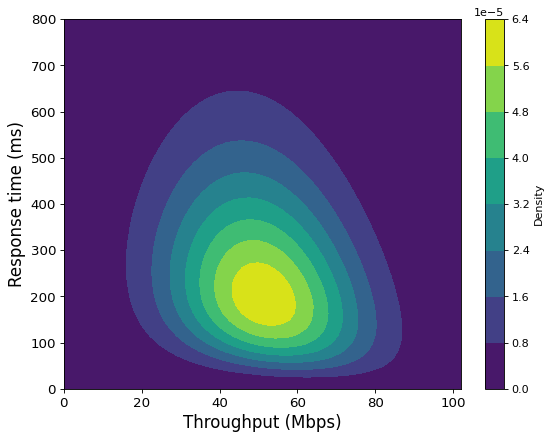}
    \caption{Depiction of a QoS profile where throughput and response time attributes are negatively correlated.}
    \label{fig:correlated-example}
\end{figure}

Notice that, as it was shown in \cref{fig:independent-example}, and formalised above, the behaviour of TP is independent from the behaviour of RT. In general, this is not the case; it is well-known that, in practice, TP and RT have a negative correlation (i.e., high RT implies low TP, and viceversa). Those situations necessarily require an approach capable of dealing with a composite multivariate probability distribution. In \cref{fig:correlated-example} we depict an example where TP and RT follow the distributions described in \cref{fig:independent-example}, but they are negatively correlated.
In this case we have a bivariate distribution $\randvec_\text{corr} = (\randvar_1, \randvar_2)$, where $\randvar_1\ \sim \mathit{Gaussian}(\mu, \sigma^2)$ and $\randvar_2\ \big\vert\ \randvar_1 \sim \mathit{Gamma}(\alpha - \frac{\randvar_1 - \mu}{\mu}, \beta)$, with $\mu = 50, \sigma^2 = 300, \alpha = 3, \beta = \frac{1}{100}$. Notice that the first parameter of the \emph{Gamma} distribution, that governs the mean of $\randvar_2$, is negatively correlated to the outcome of $\randvar_1$.

Summarising, a QoS profile is modeled by random vector whose dimension coincides with the number (and order) of the prefixed list of attributes $\qosattrset$.

\subsection{Probabilistic QoS requirements of a service client}\label{sec:qosreq}
When a QoS profile is characterised by a joint PDF it is possible to compute the probability of the execution of a service to adopt values for the quantitative attributes falling in a specific region, by integrating the joint PDF over that region. This supports the definition of probabilistic QoS requirements, whose satisfaction by a QoS profile, can be determined by queries of this type. We first introduce basic probabilistic QoS constraints.

\begin{definition}[Probabilistic QoS constraint]
\label{def:qos-constraint}
A \emph{probabilistic QoS constraint}, over the set of attributes $\qosattrset$, is a tuple containing:
\begin{inparaenum}[1)] 
\item a region of values $\region \subseteq \mathbb{R}^n$, 
\item a lower bound $\pmin \in [0, 1]$, and 
\item an upper bound $\pmax \in [0, 1]$. 
\end{inparaenum}

The set of QoS constraints over $\qosattrset$ will be denoted $\qoscstrset (\qosattrset)$.
\end{definition}

\begin{figure}[h]
    \centering
    \includegraphics[width=.5\textwidth]{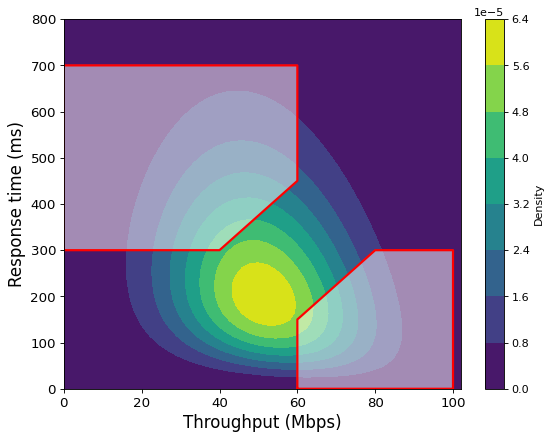}
    \caption{Regions highlighted in red depict low quality TP with low quality RT, and high quality TP with high quality RT, resp.}
    \label{fig:constraint}
\end{figure}

The intuition is that $\region$ characterizes a region of QoS values of interest, where each point $\vect \in \region$ represents a specific combination of values of the quantitative attributes, and $\pmin$ and $\pmax$ the bounds for the probability of the quantitative attributes taking values in $\region$. Then, we can consider a scenario like the one shown in \cref{fig:constraint}, where a service client is interested in a service with high quality TP and high quality RT. 

The rightmost region of \cref{fig:constraint} shows a region $R_\text{good}$ of interest for the client. $R_\text{good}$ imposes high TP (between 60 and 100 Mbps), quick RT (below 300 ms) and, additionally, excludes the lower quality section within those values (the diagonal line excludes the slow RT and low TP upper left corner). Expressing the requirement that the region of values must have high probability requires defining $\pmin = 0.8$ and $\pmax = 1$. Similarly, the leftmost region $R_\text{bad}$ characterises a poor quality TP and poor quality RT that the client might be interested in avoiding by setting a low upper bound for its expected probability.

Next, we define satisfaction of a constraint by a QoS profile.

\begin{definition}[Satisfaction of probabilistic QoS constraint]
\label{def:sat-atomic}
Given a QoS profile $\randvec$ and a QoS constraint $\qoscstr = \langle \region,\ \pmin,\ \pmax\rangle$ we say that $\randvec$ satisfies $\qoscstr$ (denoted $\randvec \models \qoscstr$) if and only if $\pmin \leq \Prob(\randvec \in \region) \leq \pmax$.
If the QoS profile $\randvec$ is described in terms of a joint PDF $f_\randvec$, then the satisfaction problem $\randvec \models \qoscstr$ is equivalent to deciding whether
    $
    \pmin \leq \int_\region f_\randvec(\randobs_1, \ldots, \randobs_n)\ d\randobs_1 \ldots d\randobs_n \leq \pmax
    $
\end{definition}

Furthermore, we can define a more expressive language for characterising QoS requirements, together with an appropriate notion of satisfaction by a QoS profile.

\begin{definition}[Probabilistic QoS requirement]
    \label{def:qos-requirement}
Given a set of propositional variables $\propset$, a \emph{probabilistic QoS requirement} (denoted by $\qosreqset (\qosattrset, \propset)$) is a formula in the language of the next grammar,
  $
    \qosreq ::= \top\ |\ \bot\ |\ \propvar\ |\ \qoscstr\ |\ \neg \qosreq\ |\ \qosreq \lor \qosreq
  $
\noindent where $\qoscstr \in \qoscstrset (\qosattrset)$ and $\propvar \in \propset$.
\end{definition}

The intuition is that QoS requirements are Boolean combinations of QoS constraints and pure Boolean variables. While it is natural for QoS requirement to be built from QoS constraints because they are the key element in modeling QoS behaviour, pure Boolean variables play an important role in facilitating the task of contract design expressing the discrete aspects of the models. 
The following QoS requirement $\qosreq_1$ expresses a relatively high probability for the values to fall in $R_\text{good}$ and a low probability of them to fall in $R_\text{bad}$,
    $\langle R_\text{good},\ \pmin: 0.6 \rangle \land \langle R_\text{bad},\ \pmax: 0.3 \rangle$.
A client might be willing to accept any of the two scenarios, as far as it can be identified which one was satisfied. This can be done by resorting to Boolean variables as follows
\begin{gather*}
    (\langle R_\text{good},\ \pmin: 0.6 \rangle \lor \langle R_\text{bad},\ \pmax: 0.3 \rangle) \\
    \land\ (p_1 \Leftrightarrow \langle R_\text{good},\ \pmin: 0.6 \rangle) 
    \land (p_2 \Leftrightarrow \langle R_\text{bad},\ \pmax: 0.3 \rangle)
\end{gather*}
In this way, we bound the truth value of each propositional variable to the satisfaction of each QoS constraint. Propositional variables $p_1, p_2$ can then be used as decision variables in another fragment of the same formula, or they could be read by the client if the verification procedure returns a satisfying assignment.

The next definition extends the notion of satisfaction of QoS constraints by a QoS profile to QoS requirements.

\begin{definition}[Satisfaction of probabilistic QoS requirement]
    \label{def:satisfaction}
A probabilistic QoS profile $\randvec$ satisfies a probabilistic QoS requirement $\qosreq \in \qosreqset (\qosattrset, \propset)$ if and only if there exists a valuation $\propval : \propset \to \{\top, \bot\}$ (i.e. a valuation of the propositional variables) such that $\randvec, \propval \models \qosreq$. With the entailment relation $\models$ defined as follows:
    $$
    \begin{array}{rcll}
        \randvec, \propval & \models & \top & \text{ holds }\\
        \randvec, \propval & \models & \bot & \text{ does not hold }\\
        \randvec, \propval & \models & p &  \text{ iff }  \propval(p) = \top \\
        \randvec, \propval & \models & \qoscstr & \text{ iff } \randvec \models \qoscstr \text{ (\cref{def:sat-atomic}),}\\
        & & & \text{ where } \qoscstr \in \qoscstrset (\qosattrset). \\
        \randvec, \propval & \models & \neg\qosreq & \text{ iff } \randvec, \propval \models \qosreq \text{ does not hold }\\
        \randvec, \propval & \models & \qosreq \lor \qosreq' & \text{ iff } \randvec, v \models \qosreq \text{ or } \randvec, \propval \models \qosreq'\\
        & & & \text{ where } \qosreq, \qosreq' \in  \qosreqset (\qosattrset, \propset). 
     \end{array}
    $$
\end{definition}

Consider the first QoS requirement example $\qosreq_1$ defined above and the QoS profile $\randvec_\text{corr}$ of \cref{fig:correlated-example}. Under this definition, $\randvec_\text{corr} \models \qosreq_1$ does not hold because, while $\int_{\region_\text{bad}} f_{\randvec_\text{corr}} \approx 0.24$ satisfies the second constraint, $\int_{\region_\text{good}} f_{\randvec_\text{corr}} \approx 0.13$ does not satisfy the first constraint, rendering the conjunction unsatisfiable.

\section{Automatic analysis of QoS requirements}
\label{sec:implementation}
Given our framing of the problem, introduced in \cref{sec:intro}, an algorithm for deciding the satisfaction relation of \cref{def:satisfaction} would effectively solve the service selection problem, by giving a decision criteria for each service in the repository. In this section we give such algorithm, which has two components: the problem of numerically integrating joint PDFs to determine the satisfaction of QoS constraints, and the algorithmic satisfaction of the Boolean structure of the QoS requirement.
We start by defining a simple transformation of QoS requirements that will enable the use of state-of-art SAT-solvers for efficiently dealing with the Boolean structure of the formula.

\begin{definition}[Propositional abstraction of QoS requirement]
    \label{def:qos-requirement-abs}
    Let $\qosattrset$ be a set of quantitative attributes, $\propset$ a set of Boolean variables and $\qosreq \in \qosreqset (\qosattrset, \propset)$ a probabilistic QoS requirement. Now, assume that $C$ is the set of probabilistic QoS constraints occurring in $\qosreq$. We define the \emph{propositional abstraction} of $\qosreq$ as a Boolean formula $\qosreq_B$ obtained from $\qosreq$ by replacing each probabilistic QoS constraint with a fresh propositional variable $\propvarb_i, i \in \{ 1, \ldots, |C| \}$.
\end{definition}

\Cref{alg:qoscheck} describes the procedure for deciding whether $\qosprf \models \qosreq$ holds or not. Essentially, there are three stages:
\begin{inparaenum}[1)]
\item it builds the propositional abstraction of $\qosreq$, 
\item it evaluates each QoS constraint in order to decide the truth value of the corresponding propositional variable, and
\item it checks if the resulting Boolean formula is satisfiable. 
\end{inparaenum}
The algorithm relies on two external procedures: \textsc{SAT} and \textsc{ComputeIntegral}. There exist efficient industrial-strength implementations of the former \cite{een:sat04,gebser:lpnmr07,leberre:jsbmc-7_2-3} that can be easily employed, as off-the-shelf components, in order to solve \textsc{SAT}$(\qosreq_B)$, the satisfiability problem for the propositional abstraction of the requirement .

\begin{algorithm}
    \caption{Decision procedure for QoS requirements}
    \label{alg:qoscheck}
    \SetAlgoLined
    \SetKwFunction{qoscheck}{\textsc{QoSCheck}}
    \SetKwFunction{satcheck}{\textsc{SAT}}
    \SetKwFunction{integral}{\textsc{ComputeIntegral}}
    \KwData{QoS profile density function $f_\qosprf$, QoS requirement $\qosreq$.}
    \KwResult{Boolean indicating whether $\qosprf \models \qosreq$ holds or not.}
     Let $C$ be the set of QoS constraints that appear in $\qosreq$.\\
     Let $d$ be a dictionary from propositional variables to QoS constraints.\\
     $\qosreq_B \leftarrow \qosreq$\\
     \ForEach{QoS constraint $\qoscstr$ in $C$}{
         Let $c$ be a fresh propositional variable.\\
         Replace $\qoscstr$ with $c$ in $\qosreq_B$.\\
         $d[c] \leftarrow \qoscstr$
     }
    \ForEach{key $c$ in dictionary $d$}{
        Let $\langle \region,\ \pmin,\ \pmax\rangle$ be the QoS constraint $d[c]$.\\
        \eIf{$\pmin \leq $ \integral{$f_\randvec, \region$} $ \leq \pmax$}{
            $\qosreq_B \leftarrow \qosreq_B \land c$
        }{
            $\qosreq_B \leftarrow \qosreq_B \land \neg c$
        }
    }
    \Return{\satcheck{$\qosreq_B$}}
\end{algorithm}

Moreover, given a joint PDF $f_\randvec$ and $\region \subseteq \mathbb{R}^n$, the procedure \textsc{ComputeIntegral}$(f_\randvec, \region)$ for computing $\int_\region f_\randvec$ needs a more thorough discussion. 

\subsection{Numerical integration of QoS profile $f_\randvec$ in region $\region$}
Notice that, since the integration limits are already specified by $\region$, we are not interested in the problem of indefinite or symbolic integration \cite{bronstein05}. In contrast, the problem we face is the numerical evaluation of a definite integral \cite{davis84}. In general, the algorithms for numerically integrating a given function are classified by the type of functions and the type of regions that can be integrated. 

In our approach, functions are arbitrary PDFs (usually referred to as non-parametric PDFs) emerging from the probabilistic modeling of the QoS profile (the behaviour of the quantitative attributes of a service). Based on such a broad interpretation of quantitative attribute, we disregard those integration methods designed for specific classes of functions, such as the ones tailored for polynomials \cite{deloera:cg13}. Secondly, regions result from the constraints restricting the valuations for the quantitative attributes, leading to the problem of choosing a language for describing constraints. This requires considering two important aspects; on the one hand, the expressivity of the language, relativised by its usefulness (both from the point of view of the language usability, and from the practical relevance of the regions it can describe) and, on the other hand, the existence of efficient algorithms for solving the numerical integration. 

The most basic approach to numerical integration in multiple dimensions is to find a way of reducing it to an iteration of one-dimensional integrals, which usually assumes the region $\region$ is an $n$-dimensional rectangle \cite{vandooren:jcam-2_3}. The integration over an $n$-dimensional simplex (see \cref{def:simplex}) has been also thoroughly studied \cite{grundmann:sjna-15_2}. From these building blocks, a plethora of methods have been developed for more general regions that can be decomposed into those elementary ones \cite[Chap. 5.5]{davis84}. This is due to the fact that the value of the integral over the union of disjoint regions is equal to the sum of the values of the integrals over each region. In this context, convex polyhedra are probably the most studied type of regions \cite{bernardini:cd91,bueler00,ziegler95}, and have been used for many purposes in the field of software verification \cite{bagnara:tcs-410_46}. Convex polyhedra (see \cref{def:region}) are appealing because they provide a way to outline a set of values of interest and, additionally, non-convex regions can be decomposed into a union of disjoint convex ones. These observations lead us to choose disjoint and finite family of bounded polytopes\footnote{The reader should note that this approach also admits the use of non disjoint regions polyhedra, as they can be decomposed into a family of disjoint polytopes through the use of the the \emph{Principle of inclusion-exclusion} \cite[Chap.~4]{liu68}.
} for characterizing the regions for defining QoS constraints (see \cref{def:qos-requirement}).

\begin{definition}[Polyhedral region]
    \label{def:region}
    A \emph{polyhedral region} (or \emph{polytope}) $\region$ in $\mathbb{R}^n$ is a set of points $\vect \in \mathbb{R}^n$ that satisfy a system of linear inequalities $\Set{\vect}{A\vect \leq \mathbf{b}}$, where $A$ is a matrix in $\mathbb{R}^{m \times n}$ and $\mathbf{b} \in \mathbb{R}^{m}$.
\end{definition} 

\begin{definition}[Simplex]
    \label{def:simplex}
    Let $\{\vect[0], \ldots, \vect[n]\}$ be a set of affinely independent points in $\mathbb{R}^n$, the $n$-simplex yield by $\vect[0], \ldots, \vect[n]$ is the set of all points $\vect \in \mathbb{R}^n$ such that
    $\vect = \sum_{i=0}^n t_i \vect[i], \text{ for } t_i \geq 0 \text{ where } \sum_{i=0}^n t_i = 1$.
\end{definition}

There exist many different definitions for polytope \cite{ziegler95}, we adopted the one based on intersecting half-spaces. When an intersection polytope is bounded, we have a \emph{bounded polytope} that, as stated in \cite[Thm. 2.15]{ziegler95}, can be equivalently represented as a set of simplices that have no interior points in common. The process of building this simplicial representation from a polytope is called \emph{triangulation} or \emph{decomposition} and there are several algorithms for achieving it \cite[Sec.~3.1]{gritzmann94} (alternatively \cite{bueler00}). This decomposition completes a procedure for numerically integrating functions over regions characterised as bounded polytope. The complexity of the procedure is primarily dominated by the amount of simplices obtained by the decomposition, because all of them have to be integrated separately. The number of simplices resulting from the decomposition of a bounded polytope largely depends on the specific geometry of the polytope being decomposed reaching magnitude in $O(2^n)$ or $O(n!)$, where $n$ is the number of dimensions (in our context, the number of quantitative attributes). If the number of number of quantitative attributes is small the approach is acceptable, but if it's large the complexity threatens its applicability.

This limitation can be overcome by computing the integrals by resorting to a Monte Carlo approximation algorithm \cite{evans00}, which we will focus on for the rest of the paper. For that to be possible the following operations must be realizable:
\begin{inparaenum}[a)]
    \item compute $V$, the volume of $\region$,
    \item uniformly sample $k$ points $\{ \vect[1], \ldots \vect[k] \} \in \region$, and
    \item given a point $\vect \in \mathbb{R}^n$, compute $f_\randvec(\vect)$.
\end{inparaenum}
Then, we can compute an estimator of the integral, $Q_k = V \frac{1}{k} \sum_{i=1}^k f_\randvec(\vect[i])$ which, by the strong law of large numbers \cite[Chap.~2]{durrett19}, is known to be an approximation of the value of the integral, in the sense that $\Prob(\lim_{k \to \infty} Q_k = \int_\region f_\randvec) = 1$. The interested reader should review \cite[Sec.~6.2]{evans00} for a general presentation of this technique.

Let us review the satisfaction of the aforementioned assumptions for the Monte Carlo approximation algorithm:
\begin{itemize}[a)]
\item \textbf{Computation of the volume V:} the complexity of computing the exact volume of a bounded polytope is proven to be \textbf{\#P-Hard} \cite{dyer:sjc-17_5}, but practical implementations can be found in \cite{bueler00}. Nevertheless, efficient estimation algorithms have been developed, such as the one presented in \cite{ge:faw15}. This estimation algorithm is based on a \emph{Multiphase Monte-Carlo algorithm} and is shown to have an algorithmic complexity of $O^*(m n^3)$, where $n$ is the dimension and $m$ is the number of facets (i.e., the number of linear inequalities)\footnote{$O^*$, known as Soft-O notation, indicates the omission of logarithmic factors as well as factors depending on other parameters like the error bound.}.
\item \textbf{Uniform sampling of points in the polyhedral region:} several strategies could be used, for example, \emph{Markov Chain Monte Carlo algorithms} like \emph{Dikin walk}, introduced in \cite{kannan:mor-37_1}, with algorithimc complexity in $O(mn)$, where $n$ is the dimension of the polytope and $m$ the number of constraints; newer approaches \cite{chen:jmlr-19_1}, also based in MCMC algorithms, present different sampling strategies with better complexity under specific conditions, for example, the \emph{Vaidya walk} has complexity in $O(m^{0.5}n^{1.5})$, which is significantly better when $m \gg n$.
\item \textbf{Evaluation of the density function:} we assume throughout this work that the density function $f_\randvec$ has a finite representation, such as an \emph{algebraic expression} or a \emph{probabilistic program}, that yields precise computational steps.
\end{itemize}
Under these conditions, by \cite[Thm.~6.1]{evans00}, the error $|Q_k - \int_\region f_\randvec|$ is of order $1/\sqrt{k}$. The factor $\sqrt{k}$ comes from the distribution given by the central limit theorem applied to the estimator $Q_k$. The fact that this convergence rate does not depend on the dimension $n$ is one of the advantages of the method over other numerical methods for integration.

%% file: learning.tex

\section{Learning QoS profiles from data}\label{sec:learning}
A significant part of the literature about service-based software systems is devoted to monitoring the services' runtime behaviour and almost always such information is used to address the problem of contract violations. This standpoint can be thought of as a Foucaultian\footnote{The term is used by analogy to Michel Foucault's theorisations on how modern society deals with non-compliance with the social contract, introduced in his 1975 book ``Surveiller et punir: Naissance de la prison''.} approach in which service providers are trusted to perform according to the contract they expose, and punished if they incur in a violation. We commit to an opposite view that can be thought of as an Aristotelian\footnote{The term is used in relation to the famous Aristotle's quote ``The only truth is reality'', phrase in which he distance himself from the idealism in Plato's  philosophy.} approach; in it, the QoS profile of a service (i.e., the joint PDF of the values taken by its quantitative attributes) should be the result of learning from its behaviour in the wild, by monitoring the attributes at runtime.

Given the set of quantitative attributes $\qosattrset$ of a service, a \emph{QoS record} will be a vector $\vect \in \mathbb{R}^n$. The intuition is that a QoS record formalises the values taken by the quantitative attribute in a single execution of the service, and the idea is to estimate the QoS profile of a service by collecting its successive QoS records and estimating a joint PDF from them. In the field of probability and statistics this is a well-known problem called \emph{probability density estimation} \cite{silverman98} and there are several methods for estimating probability density functions from a sample of the underlying distribution. We illustrate the approach by resorting to the well-know technique of \emph{kernel density estimation} \cite{fix:isr-57_3,rosenblatt:ams-27_3}.

\begin{definition}[Kernel density estimation]
    Given a set of observations \\$\{\vect_1, \ldots, \vect_m\}$ in $\mathbb{R}^n$, the kernel density estimator (KDE) $\hat{f}(\vect)$ is defined as:
    $$
    \hat{f}(\vect) = \sum_{i=1}^m K_h (\vect - \vect_i)
    $$
    \noindent where $K_h$ is a kernel function satisfying $\int_{\mathbb{R}^n} K_h(\vect)\ d\vect = 1$, with $h$ being the \emph{bandwidth} or \emph{smoothing parameter}. 
\end{definition}

There are many types of kernel functions that can be used in KDE, and each of them represent a different interpretation of how the influence (weight) associated to an observation is distributed in its neightborhood. 
%
%
Once that kernel is fixed, the bandwidth plays a crucial role in how that KDE learns the distribution from the observations, if the bandwidth is too low the KDE will overfit the observations and be too influenced by the concrete values and if it is too high, it will not be able to learn and adjust to the weight of the observations. While there is no way to know the optimal bandwidth in advance there exists several strategies for identifying good candidates. The fastest methods for computing the approximate bandwidth are known as rules of thumb. Examples of such rules for Gaussian kernels are the \emph{Silverman's rule of thumb} \cite[pp.~48,~eq.~3.31]{silverman86}
, for obtaining a good bandwidth for univariate KDEs; a multivariate version of this rule \cite[pp.~152,~eq.~6.41]{scott92}
; and the \emph{Scott's rule of thumb} \cite[pp.~152,~eq.~6.42]{scott92}. 
For other kernels, \cite[table~6.3]{scott92} provides constants by which $h_i$ must be multiplied.
Full cross-validation \cite{sain:jasa-89_427} is a technique for obtaining an optimal combination of a kernel type and a bandwidth from available data. This method is significantly more computationally demanding but their use is acceptable if the parameter update is performed selectively. More sophisticated kernel estimators have been developed lately centered in the idea of incremental density estimation for streams \cite{he:complexity20}. 

In \cref{tab:kernels} we show a comparison between the approximate volume computation of the regions depicted in \cref{fig:constraint} over the exact QoS profile of \cref{fig:correlated-example}, and over KDEs trained with different sample sizes, based on optimal combinations of kernel and bandwidth, obtained by full cross-validation. 

\begin{table}[h]
\begin{center}
\begin{tabular}{|r|r|r|r|r|r|}
\hline
\multicolumn{1}{|c|}{\cellcolor{gray!50}KDE sample size} & \multicolumn{1}{|c|}{\cellcolor{gray!50}(kernel, bandwidth, time)} & \multicolumn{2}{|c|}{\cellcolor{gray!50}Region $R_\text{good}$} & \multicolumn{2}{|c|}{\cellcolor{gray!50}Region $R_\text{bad}$} \\
\hline
\hline
\cellcolor{gray!25}100 & (exponential, 13.107, 41.1s) & \multicolumn{1}{|r}{0.1729} & \multicolumn{1}{r|}{(23.57\%)} & \multicolumn{1}{|r}{0.2021} & \multicolumn{1}{r|}{(11.45\%)}\\
\hline
\cellcolor{gray!25}1000 & (exponential, 6.903, 7m 59,3s) & \multicolumn{1}{|r}{0.1546} & \multicolumn{1}{r|}{(10.49\%)} & \multicolumn{1}{|r}{0.2694} &\multicolumn{1}{r|}{(18.03\%)} \\
\hline
\cellcolor{gray!25}10000 & (exponential, 4.145, 135m 4.1s) & \multicolumn{1}{|r}{0.1312} & \multicolumn{1}{r|}{(6.20\%)} & \multicolumn{1}{|r}{0.2400} & \multicolumn{1}{r|}{(5.16\%)}\\
\hline
\hline
\multicolumn{2}{|c|}{\cellcolor{gray!50}exact QoS profile} & \multicolumn{2}{|r|}{0.1399} & \multicolumn{2}{|r|}{0.2282} \\
\hline
\end{tabular}
\end{center}
\caption{Comparison between the approximate volume computation of the regions of \cref{fig:constraint} using the exact PDF and KDEs.}
\label{tab:kernels}
\end{table}

\begin{figure}[h]
    \centering
    \includegraphics[width=.45\textwidth]{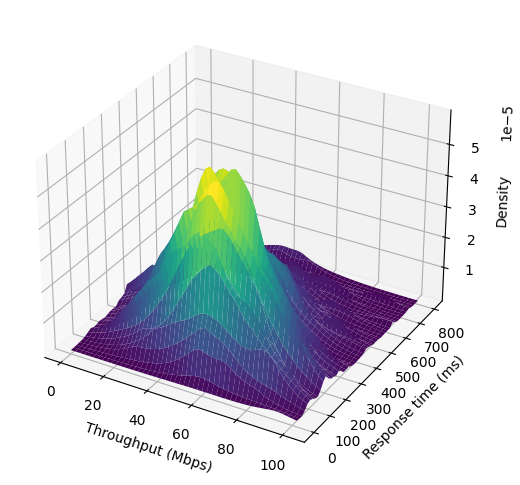}
    \includegraphics[width=.45\textwidth]{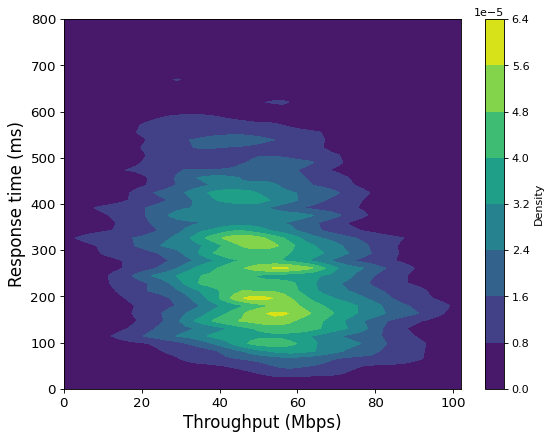}
    \caption{Depiction of a QoS profile learnt by KDE with sample size 1000 and full cross validation for kernel and bandwidth optimisation.}
    \label{fig:correlated-kde-example}
\end{figure}

In \cref{fig:correlated-kde-example} we show a plot of the joint PDF learnt by KDE from sampling 1000 observations from the exact joint PDF, presented as the running example, and resorting to full cross-validation for optimising the kernel choice and the bandwidth.

By resorting to learnt joint PDFs, QoS profiles provide the information that better fits the already observed service's behaviour, avoiding:
\begin{inparaenum}[1)]
\item requiring the service provider to give a, potentially, complex formalisation of the statistical properties of the quantitative attributes (specially considering the case of those whose behaviour fall out of the providers' control, like latency or bandwidth of the communication infrastructure) and
\item establishing complex penalisation schemes that, due to the complexity of providing a realistic joint PDF, could be unfair, or unnecessarily harsh, with a service provider.
\end{inparaenum} 

A similar idea has been studied, with promising experimental results, in \cite{zheng:soca10}. There, the authors propose the use of statistical methods to estimate the univariate PDF of each specific attribute in isolation. The contribution of this section can be thought of as a generalisation of their approach to the multivariate case, which we believe is more faithful to reality. 

It is worth noticing that QoS profiles learnt by creating KDEs might lead to unrealistic phenomena. Many times we see attributes whose PDF is not continuous, for example, a pricing scheme might be based on predefined pricing functions applicable to different consumer profiles. A KDE is not capable of showing such discontinuities. Therefore, predicting the probability using a KDE at those specific points, do not appropriately characterise the expected behaviour. While this limitation might be consider a small problem in general, it can lead to erroneous predictions, for example, when the probability associated to paying a negative price is non zero. Solving this problem requires a mayor modification of the method because we need to adapt how observations are integrated to the KDE whenever they influence non admissible values. 

%% file: conclusions.tex

\section{Conclusions}
\label{sec:conclusions}
In the present work we addressed the QoS-aware service selection problem in a probabilistic setting. We identified some elements of interest that the existing literature have treated partially and separately:
\begin{inparaenum}[1)]
\item QoS attribute are just measurable attributes not assuming any prescribed specific interpretation or behaviour, and
\item QoS attribute adopt values from a continuous probability distribution. 
\end{inparaenum}
To that we add the need for adopting a multivariate view regarding the probability distribution of the values, in order to enable the representation of interdependent probabilistic behaviour of the QoS attributes. This need was justified by resorting to the well-known existing relation between the throughput and the response time of a service (see the running example presented above).

To tackle the problem, we proposed a unified approach in which:
\begin{inparaenum}[1)]
\item QoS attributes are just real variables with no further interpretation or assumed intrinsic behaviour, 
\item QoS profiles are defined as non-parametric joint PDFs for random vectors, representing the QoS attributes, and 
\item QoS requirements are Boolean combinations of probabilistic constraints;
\end{inparaenum} 
for which we provided:
\begin{inparaenum}[1)]
\item an algorithm for deciding whether a QoS requirement is satisfied by a QoS profile, based on an efficient Monte Carlo integration procedure and state-of-the-art SAT-solving techniques, and
\item the mathematical foundations for learning QoS profiles from empirical data sampled from executions, eliminating the burden of creating QoS profiles from the service providers. 
\end{inparaenum}
We also discussed how the presented algorithm constitutes the core procedure for solving the service selection problem in a probabilistic environment.
